\documentclass[a4paper,11pt]{article}
\pdfoutput=1
\usepackage[utf8x]{inputenc}
\usepackage{jheppub}
 
\usepackage{amsmath,amssymb}
\usepackage{subfig}
\usepackage{float}
\usepackage{slashed}
\usepackage{multirow}

\title{Explanation of IceCube spectrum with $\nu\rightarrow 3 \nu$ neutrino splitting in a  $\nu$2HDM model}

\author[a]{Subhendra Mohanty}
\author[b]{Soumya Sadhukhan}

\affiliation[a]{Physical Research Laboratory, Ahmedabad 380009, India}
\affiliation[b]{Physical Research Laboratory, Ahmedabad 380009, India} 

\emailAdd{soumyas@prl.res.in}
\emailAdd{mohanty@prl.res.in}

\vspace{1cm}

\abstract{A single power law flux spectrum of high energy neutrinos does not adequately explain the entire 60 TeV  to 10 PeV event spectrum observed at IceCube, 
specially the excess of PeV events and the lack of Glashow resonance events expected at 6.3 PeV cannot be simultaneously explained by a single power law source neutrino flux.
Here we consider a model of neutrino splitting $\nu\rightarrow 3 \nu$ over cosmological distances. 
Starting from a single power-law spectrum expected from the astrophysical sources, 
we show that by adjusting the decay length and spectral index one can give a better fit to the observed IceCube events over the entire 1 TeV -6 PeV, 
compared to that from a single power spectrum.
For $\nu\rightarrow 3\nu$ neutrino splitting, the flavor ratios of the daughter neutrinos are different from the standard oscillation or invisible decay cases and
can be used as a test of this scenario.
We propose a $\nu$2HDM where a light Higgs ($\sim 0.1$~eV) mediates neutrino splitting via a one-loop box diagram. The split in the masses of the scalars in the doublet gives a large contribution to the oblique T parameter which
is severely constrained. This constraint from the S,T,U oblique parameters can be evaded by the introduction of an extra vector lepton doublet and  with mass $\sim 200$ GeV.}

\allowdisplaybreaks

\begin{document}
\maketitle
\flushbottom

\section{Introduction}
\label{sec:introduction}

After seven years of IceCube measurement, a clear 6$\sigma$ excess of events is observed for energies above 60 TeV which cannot be 
explained by the atmospheric neutrinos~\cite{Aartsen:2017mau, Aartsen:2015rwa}. 
The first choice for explaining the high energetic neutrino events were different astrophyscal sources~\cite{Cholis:2012kq, Anchordoqui:2013dnh, Murase:2014tsa, Sahu:2014fua}. 
The source and spectrum of neutrinos observed at IceCube remains a puzzle
as the events do not point back to any clear identifiable Active Galactic Nuclei (AGN) or Gamma Ray Bursts (GRB) and the entire spectrum of events from 60 TeV to 10 PeV cannot be explained by a single power law of neutrino flux like $\Phi_\nu= \Phi_0 E_\nu^\alpha$. 
In GRB models ~\cite{Waxman:1998yy}, charged pions are produced when high energetic protons interact with gamma background as 
$p + \gamma \to \Delta^{+} \to n + \pi^{+}$, followed by further decay of pions as $\pi^{+} \to \mu^{+} \nu_{\mu} $ and $\pi^{+} \to e^+ \nu_{e} \nu_{\mu}$, 
producing flux of high energetic muon and electron neutrinos. IceCube is sensitive to those neutrinos whose energies are above 0.1 TeV. 
In general the energy dependence of the incoming neutrino flux from the GRBs 
at the source is given by the Waxman Bahcall spectrum~\cite{Waxman:1998yy} $\Phi_\nu \propto  E^{-2}$. 
The astrophysical neutrino flux modeled in this way does not give a good fit to the IceCube event distribution at all energy bins upto $ \sim 2 $~PeV. 
With a fit at lower energy ($\sim 100$~TeV) bins, this model predicts an excess of neutrino events at higher energy bins, but IceCube has not observed that effect till now. 
Even if the decrease of flux amplitude can fit neutrino event observation at some bin, there remains huge mismatch with predicted events at other energies. 
If a steeper neutrino energy spectrum is taken, IceCube event distribution in the sub-PeV bins can be explained by fixing a proper neutrino flux amplitude. 
The steepness of the spectrum will result in deficiency of predicted events at energies $1-3$~PeV. Invisible and visible decays of astrophysical 
neutrinos is explored in this work as a better explanation of 6 years of IceCube event distribution data. Various models for explaining the observed features in the IceCube event spectrum include neutrinos from Pev dark matter decay or annihilation to explain the 1 PeV excess (\cite{Esmaili:2013gha, Murase:2015gea, Dev:2016qbd, Bhattacharya:2014yha}). 
The excess of PeV events can also come from the resonant production of leptoquarks ~\cite{Barger:2013pla, Dutta:2015dka, Dey:2015eaa, Chauhan:2017ndd}. 
On the other hand, there are neutrino depletion models which try to explain the non-observation of the Glashow resonance ~\cite{Glashow:1960zz,Berezinsky:1977sf,Berezinsky:1981bt} i.e. absence of the resonant peak due to a real $W^-$ production in the process ${\bar \nu}_e e^- \to W^-$. 
The decay of the real $W$ is expected to give hadron and lepton shower or lepton track events ~\cite{Bhattacharya:2011qu}. Depletion of high-energy neutrinos can occur via oscillation to sterile neutrinos in pseudo-Dirac neutrinos~\cite{Joshipura:2013yba} and for visible decay ~\cite{Pakvasa:2012db}. Exotic scenarios have also been invoked to explain a cutoff at the Glashow resonance energies such as Lorentz violation ~\cite{Tomar:2015fha, Anchordoqui:2014hua} and CPT violation~\cite{Liao:2017yuy}. 

Neutrino splitting $\nu\rightarrow 3 \nu$ can change the neutrino spectrum at IceCube and it can occur in CPT and Lorentz violation models \cite{Cohen:2011hx, Liao:2017yuy}. 
In this paper we consider a Lorentz invariant BSM model where the neutrino splitting $\nu \rightarrow 3 \nu$ occurs through a one-loop diagram with a light mediator.
One BSM case which can give rise to the neutrino decay is neutrinophilic two Higgs doublet model~\cite{Machado:2015sha} 
($\nu$2HDM) which was originally proposed to explain the nonzero but tiny neutrino mass. 
This kind of neutrinophilic models of neutrino mass generation was initially proposed and discussed in the References~\cite{Gabriel:2006ns, Davidson:2009ha}.
In $\nu$2HDM, neutrino masses are generated by introducing a second scalar doublet with a vacuum expectation value of the eV scale. 
The astrophysical neutrino decay rate should be large enough to have a decay before neutrinos reaching the IceCube detector. 
Larger decay rates are ensured in the $\nu$2HDM through the presence of a eV scale neutral scalar. 
Presence of an ultra light scalar results in a stringent bound from the oblique parameters. 
This can be resolved by expanding the model with vectorlike leptons, which themselves have interesting phenomenological properties. 

Neutrino decay to visible lower energy neutrinos change the spectrum of IceCube neutrinos and give a better fit of the IceCube event spectrum in the entire range 60 TeV -10 PeV
compared to a single power law neutrino flux. 
We note that in the neutrino splitting scenario, it protects the information of astrophysical source direction, as the daughter neutrinos are collinear being very energetic ones, which is consistent with the recent observation that one single source can be traced back from the observed IceCube spectrum.

In section~\ref{model} we have briefly explained the $\nu$2HDM model along with the explanation of emergence of tiny neutrino mass there. 
In the next section~\ref{ndecay}, it is shown how active SM neutrinos can split to three active neutrinos through the box loop diagram. We have also 
shown the presence of an eV scale scalar propagator increases the decay rate and therefore ensure a finite decay lifetime to have any astrophysical effect. 
In the next section~\ref{nusplit}, we show the flavor ratios of the daughter neutrinos after splitting, taking into account different astrophysical 
sources with different initial flavor ratios.
In section~\ref{obp}, we discuss how the presence of an ultra light scalar in $\nu$2HDM can have large negative contribution on $S, T$ parameters to push 
them to the values which are not experimentally viable. In section~\ref{oc} different flavor, collider and theoretical constraints are discussed along with 
their bounds on $\nu$2HDM parameter space. In section~\ref{icana}, we do the analysis of the effects of both the neutrino splitting ($\nu \rightarrow 3 \nu$) and invisible decays  on the 
astrophysical neutrino propagation and their event distribution at IceCube. In one subsection~\ref{vd} we concentrate on visible neutrino decay and show how its presence 
fit the IceCube event distribution along with an comparison with the case where there is no neutrino decay. In another subsection~\ref{invd}, similar studies are done for 
neutrino invisible decay. Finally we summarize and conclude in section~\ref{conc}.


 \section{The Model}
\label{model}
The beyond the SM $\nu$2HDM theory~\cite{Gabriel:2006ns, Machado:2015sha} discussed in this work, is based on the symmetry group $SU(3)_c \times SU(2)_L \times U((1)_Y \times Z_2$. 
In addition to the usual SM fermions, we have one EW singlet right-handed (RH) neutrino, $\nu_{R}$, for all flavors of SM leptons.  
The model has two Higgs doublets, $\Phi_{1}$ and $\Phi_{2}$.  
All the SM fermions and the Higgs doublet $\Phi_{1}$, are even under the discrete symmetry, $Z_2$, while
the RH neutrino and the Higgs doublet $\Phi_{2}$ are odd under $Z_2$.
This leads to Yukawa interaction of all the SM fermions except the left-handed neutrinos, through $\Phi_{1}$ only. 
The SM left-handed neutrinos, together with the right-handed neutrino added here, couple to the Higgs doublet $\Phi_{2}$. 
The discrete symmetry $Z_2$ is broken by a vev of $\Phi_{2}$, and we take $v_2 = \langle\Phi_{2}\rangle \sim 0.1$~eV. 
Therefore, the origin of the neutrino mass in $\nu$2HDM is due to a spontaneous breaking of the discrete symmetry $Z_2$.
Through their Yukawa interactions with the Higgs field $\Phi_{2}$, the neutrinos acquire orders of magnitude smaller masses 
than the SM ones even with order one Yukawa couplings, due to the smallness of $v_2$.

For simplicity, we consider the Higgs sector to be CP invariant here. 
The most general Higgs potential consistent with the $SM \times Z_2$ symmetry
is \cite{Machado:2015sha}
\begin{align}
\ V =
-\mu^2_1~\Phi_{1}^{\dag}\Phi_{1}-\mu^2_2~\Phi_{2}^{\dag}\Phi_{2}+\lambda_1(\Phi_{1}^{\dag}\Phi_{1})^{2}+\lambda_2(\Phi_{2}^{\dag}\Phi_{2})^{2}+ 
\lambda_3(\Phi_{1}^{\dag}\Phi_{1})(\Phi_{2}^{\dag}\Phi_{2})-
\lambda_4|\Phi_{1}^{\dag}\Phi_{2}|^{2}\nonumber\\-\frac{1}{2}\lambda_5[(\Phi_{1}^{\dag}\Phi_{2})^{2}+(\Phi_{2}^{\dag}\Phi_{1})^{2}].
\label{pot}
\end{align}
In the $\nu$2HDM scalar potential of Eq.~\ref{pot}, the parameters
$m_{12}^2$ and $\lambda_{6,7}$ are put to zero due to the imposition of $Z_2$ symmetry.
The physical Higgs fields are charged fields $H^{\pm}$, two neutral CP even scalar fields $h$ and $H$, and a neutral CP odd field $A$. 
In the unitary gauge, the two doublets can be written
\begin{align}
\Phi_{1} =  \frac{1}{\sqrt{2}} \left(\begin{array}{c}
                                                 \sqrt{2} (v_2/v)H^{+} \\
                                                 h_0 + i (v_2/v)A +v_1 \\
                                               \end{array}
                                             \right),    \nonumber
\end{align}
\begin{align}
           \Phi_{2} = \frac{1}{\sqrt{2}} \left(
                                               \begin{array}{c}
                                                 -\sqrt{2} (v_1/v) H^{+} \\
                                                 H_0 - i (v_1/v) A + v_2\\
                                               \end{array}
                                             \right),
 \end{align}
 where $v_1 = \langle\Phi_{1}\rangle$, $v_2 = \langle\Phi_{2}\rangle$, and $v^{2} = v^{2}_1 + v^{2}_2$. 
Here the charged and CP odd interaction states mix with corresponding charged and neutral Goldstone modes through an orthogonal mixing with angle $\beta$. 
This mixing produces mass eigenstates $H^{\pm}, A$ along with the massless Goldstone bosons that are removed in the unitary gauge. 
The mixing angle is expressed as $\tan \beta = \frac{v_2}{v_1}$. 
The particle masses are
 \begin{align}
m^2_{H^{\pm}} = \frac{1}{2}(\lambda_4 +\lambda_5)v^{2}, ~ m^{2}_A = \lambda_5 v^{2},\nonumber
\end{align}
 \begin{align} 
m^{2}_{h,H} = (\lambda_1 v^{2}_1 +\lambda_2 v^{2}_2)\pm \sqrt{(\lambda_1 v^{2}_1 -\lambda_2 v^{2}_2)^{2}
 +(\lambda_3-\lambda_4-\lambda_5)^{2} v^{2}_1 v^{2}_2}.
 \end{align}
An immediate consequence of the scenario under consideration is a
very light scalar $H$ with mass,
\begin{align}
m^{2}_H = 2\lambda_2 v^{2}_2[1+O(v_2/v_1)]. 
\end{align}
The mass eigenstates $h, H$ are related to the weak eigenstates $h_0, H_0$ by
\begin{align}
\label{mix}
 h_0 = c_{\alpha} h + s_{\alpha} H, ~H_0 = -s_{\alpha} h + c_{\alpha} H,
  \end{align}
where $c_{\alpha} = \cos\alpha, s_{\alpha} = \sin\alpha$, and
are given by
\begin{align}
\label{scmix}
 c_{\alpha} &= 1+O(v^{2}_2/v^{2}_1),\nonumber \\
s_{\alpha} &= -\frac{\lambda_3-\lambda_4-\lambda_5}{2\lambda_1}(v_2/v_1)+O(v^{2}_2/v^{2}_1).
\end{align}
This mixing is extremely tiny as along with the absence of an explicit $Z_2$ breaking $m_{12}^2$ term, $v_{2} \sim 0.1$~eV and $v_{1} \sim 250$ GeV, 
and thus can be neglected.
The smallness of $\tan\alpha$ can be seen from Eq.~\ref{scmix}, imposing the limit $v_2/v_1\rightarrow 0$, which evidently results in tiny $\tan\beta$. 
Hence, the neutral scalar $h$ essentially behaves like the SM Higgs, except of course in
its Yukawa interaction with the neutrinos. 
So we do not expect any observable deviation from the Higgs couplings to the SM particles,
except possibly in the loop induced couplings, e.g. $h\gamma\gamma$ (through $H^{\pm}$ loop). 
The effect of $H^{\pm}$ loop in the $h\gamma\gamma$ vertex is computed in Ref.~\cite{Seto:2015rma}, 
along with the estimation that a sizable Higgs invisible decay is allowed here.
Therefore, the charged fermions obtain mass entirely from the $\Phi_1$ vev $v_1$, and neutrinos get it through its coupling exclusively to $\Phi_2$.
In the $\nu$2HDM with an imposed lepton number conservation, the neutrino mass is of Dirac nature. 
If that symmetry is not enforced, $\nu$2HDM allows Majorana mass generation of neutrinos with a low scale seesaw mechanism.  

In the $\nu$2HDM with $Z_2$ symmetry, the second Higgs doublet $\Phi_2$ shows some interesting features. 
The neutral scalars $H, A$, in their Yukawa interaction, couple almost fully to the neutrinos, while the interactions between neutrinos and 
charged leptons are mediated through $H^{\pm}$. 
Presence of tiny and nonzero neutrino mass can be explained even with the Yukawa couplings at $O(1)$ values. 
Among the $H, A$ coupling to the $W$ and $Z$ bosons, tiny $v_2$ supresses the triple gauge couplings involving only one scalar. 
The triple gauge couplings with two scalars and one gauge boson can be there and can give rise to a significant scalar pair 
production cross section at the LHC, in the processes like $pp\to {\gamma}^* \to H^+H^-$.

\section{Splitting of Active Neutrinos}
\label{ndecay}
In $\nu$2HDM scenario, $\Phi_1$ behaves like the SM scalar doublet while $h$ resembles the SM Higgs. As explained in section~\ref{model}, $\Phi_1$ couples to all the SM 
fermions except the neutrinos with the second doublet $\Phi_2$ and the scalars inside, $H,A,H^\pm$ interacting in a neutrinophilic way in the Yukawa sector. 
We add a right handed neutrino $\nu_R$, which is odd under $Z_2$ symmetry. This RH neutrino along with $\Phi_2$, which is also odd under $Z_2$ forms the Yukawa 
interaction as
\begin{align}
\label{yuk}
L_{Y} &= y_e L_{e}^{\dagger} \tilde{\Phi_2} \nu_R + y_{\mu} L_{\mu}^{\dagger} \tilde{\Phi_2} \nu_R + y_{\tau} L_{\tau}^{\dagger} \tilde{\Phi_2} \nu_R + \mathrm{h.c.}, \nonumber \\ 
&= \sum_{e, \mu, \tau} \frac{y_e v_2}{\sqrt{2}} U_{ei} \nu_i \nu_R + \sum_{e, \mu, \tau} \frac{y_e}{\sqrt{2}} U_{ei} H (i A) \nu_i \nu_R + \text{other terms} , \nonumber \\
&=  m_{\nu_i} \nu_i \nu_R + \frac{m_{\nu_i}}{v_2} H (i A) \nu_i \nu_R + \text{other terms} ,
\end{align}
where $m_{\nu_i}=\sum_{e, \mu, \tau} \frac{y_e v_2}{\sqrt{2}} U_{\ell i} $ are neutrino masses and $U_{\ell i}$ is the PMNS matrix.
%
\begin{figure}[tb]
\begin{center}
\includegraphics[width=7cm]{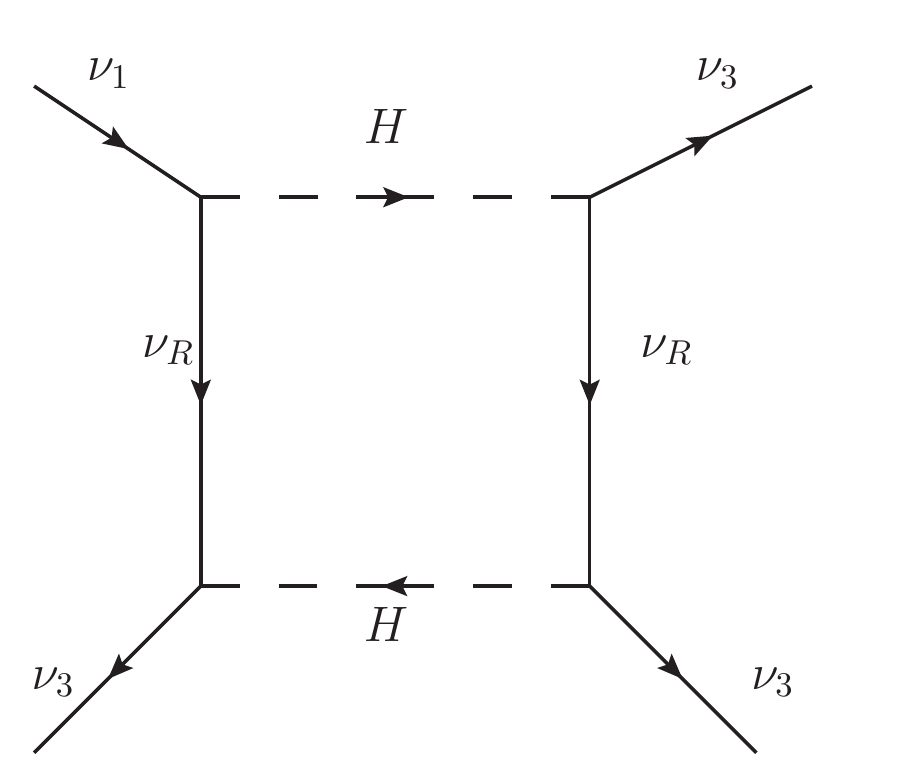}
\end{center}
 \caption{Visible decay of active neutrinos to other active neutrinos in the $\nu$2HDM set up through a box diagram: $\nu_1 \rightarrow 3 \nu_3$}
\label{ndp}
\end{figure}

We take the right handed neutrino $\nu_R$ mass to be of the order of the SM neutrino mass, i.e. $m_N \approx m_{\nu}$ and the neutrino Yukawa coupling $y_i = \frac{m_{\nu_i}}{v_2}$. 
The box diagram of neutrino decay $\nu \to 3 \nu$ is formed through the neutral scalar (H/A) and right handed neutrino propagators as shown in Fig.~\ref{ndp}. 
The $\nu_1 \to 3 \nu_3$ decay through a box diagram is computed in $\nu$2HDM and the decay rate is 
\begin{align}
\label{yu}
\Gamma_{\nu} = \frac{16 \pi^4 m_{\nu_1} m_{\nu_3}^3}{m_H^4} \ ( y_1 y_3^3 )^2 \ |F\left( \frac{m_N^2}{m_H^2} \right)|^2 \ PS_{1 \to 3},
\end{align} 
where $PS_{1 \to 3}$ is the phase space factor of one particle decay to three particles, computed following the Ref.~\cite{Romao:1998sr}. 
The loop factor $F\left( \frac{m_N^2}{m_H^2} \right)$ is obtained as 
\begin{align}
F (y) = \frac{1 + y}{(1-y)^2} + \frac{2 y \log y}{(1-y)^3}.
\end{align}
Neutrino decay lifetime $\tau_{\nu}$ is the inverse of the decay rate computed here.

In the model considered here, three light neutrinos get mass from a neutrinophilic doublet $\Phi_2$ with $y_i \sim O(1)$. 
We assume a scenario where the heaviest two of these three neutrinos can split to a number of the lightest ones, 
given that the distances they travel from their sources to the place of their detection at IceCube are large enough.  
If three neutrino masses are nearly degenerate, then the daughter neutrino carries almost all of the decaying neutrino energy, and 
thus contributing to the flux at that energy. 
So there will be no depletion in neutrino count in that energy range. 
To show the decrease in neutrino flux, the nature of the mass spectrum must be hierarchical, i.e. having two mass eigenstates with mass much higher than the final state neutrino. 
In this case, the daughter neutrinos after the splitting will be much less energetic than the parent one and therefore, those are left out of the neutrino flux at the energy scale of the splitted neutrino~\cite{Pakvasa:2012db}.
%
\begin{figure}[tb]
 \begin{center}
   \includegraphics[width=6.5cm]{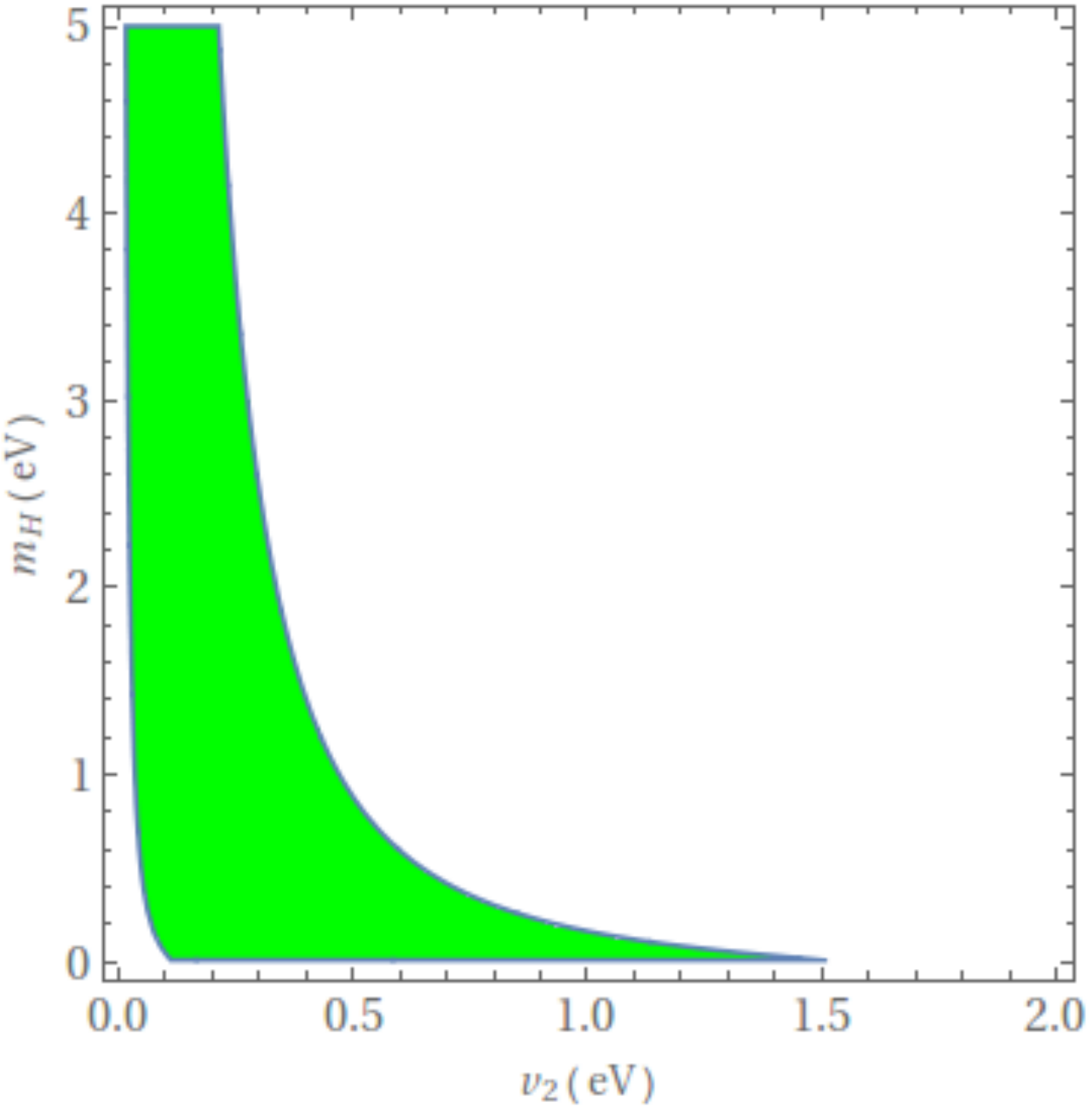}\hspace{0.3cm}
\includegraphics[width=7cm]{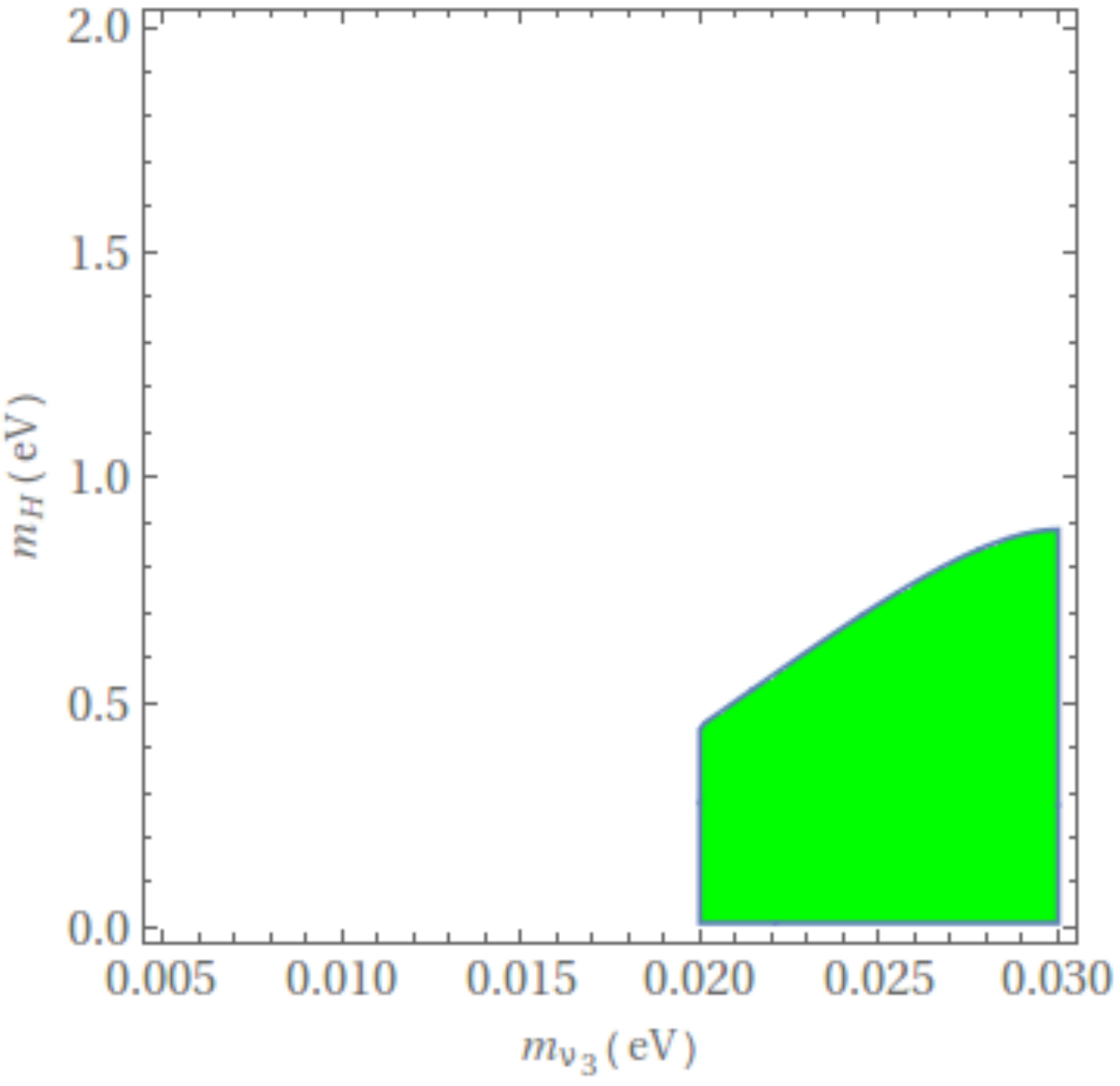}
 \end{center}
 \caption{Parameter region satisfying the allowed range of $\frac{\tau_{\nu}}{m_{\nu}}$ as shown in Eq.~\ref{range} for active neutrino decay $\nu_1 \to 3 \nu_3$, 
a. in $v_2 - m_H$ plane with $m_{\nu_3}=0.03$~eV, $m_{\nu_1}=0.1$~eV (left) b. in $m_{\nu_3} - m_H$ plane with $v_2 =0.5$~eV, $m_{\nu_1}=0.1$~eV (right).}
\end{figure}
In this work we take $\nu_3$ to be the lightest final state neutrino. The $\nu_3- \nu_e$ mixing through tiny PMNS element $U_{e3}$, suppresses the $\nu_e$ number 
that come out from a $\nu_3$ state, making the $\nu_e$ count even smaller. For $\nu_3$ to be the lightest neutrino, inverse neutrino hierarchy is preferred 
where $\nu_1, \nu_2$ are relatively heavy ($\sim 0.1$~eV) and almost with similar mass and the $\nu_3$ is extremely light ($\sim 0.01$~eV). 
 
The distances of the neutrino sources like AGN and GRB's are of order of 10's of MPc from the earth, where we hope to have the detection. 
For the final state neutrinos to be counted out, the decay factor $e^{-L/\gamma c \tau}$ should be negligibly small which translates to $\frac{L}{\gamma c \tau} \gg 1$.
For the neutrinos coming with PeV scale energies,
\begin{align}
 \frac{L}{\gamma c \tau} = \frac{L}{E} \left (\frac{mc^2}{c \tau} \right ) \gg 1, \nonumber
\end{align} 
which corresponds to $\tau/m < 10^{3} s/eV$ where $\tau$ is the neutrino rest frame lifetime.
Big Bang Nucleosynthesis (BBN) provides a lower bound on the neutrino lifetime~\cite{Pakvasa:2012db}. 
With these limits the allowed window of neutrino decay lifetime is constrained to the range~\cite{Pakvasa:2012db},
\begin{equation}
 \label{range}
10^{-6} s/eV \leq \frac{\tau}{m}\leq 10^3  s/eV~.
\end{equation}

\begin{figure}[tb]
 \begin{center}
    \includegraphics[width=7cm]{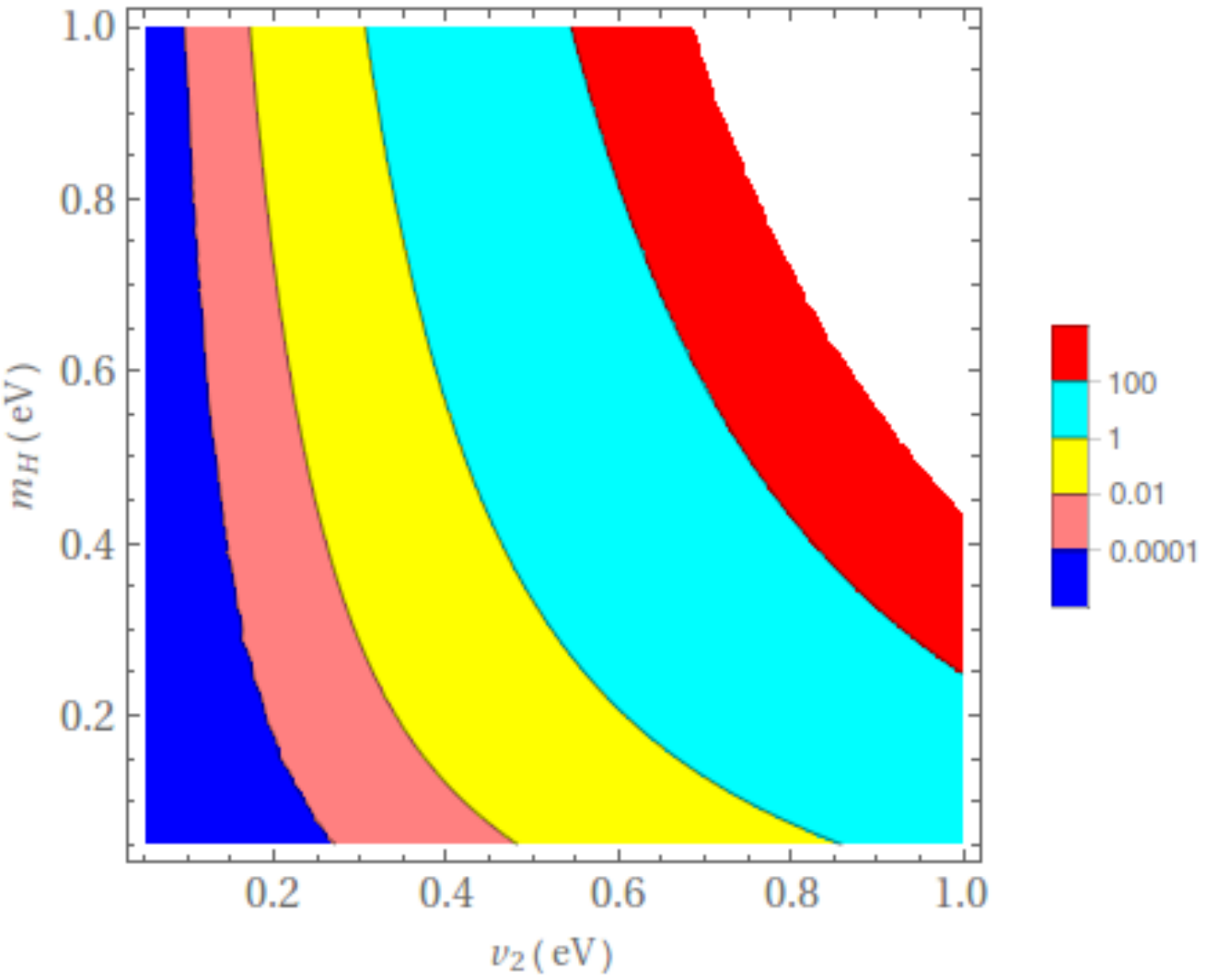}
 \end{center}
 \caption{
Different contours of allowed values of active neutrino lifetime $\tau_{\nu}$ from the $\nu_1 \to 3 \nu_3$ decay.}
\end{figure}

\section{Flavor ratios in Neutrino splitting}
\label{nusplit}
 We consider three different production mechanism of astrophysical neutrinos here and explore the effects of neutrino splitting on the flavor composition of final 
state neutrinos, both for normal and inverted mass hierarchies. 
The allowed ranges of neutrino flavor ratios at the Earth for different standard and non-standard cases, with arbitrary initial flavor ratios, is 
discussed previously in Ref.~\cite{Bustamante:2015waa}.
Flavor composition of neutrinos after neutrino oscillation and decay is discussed in Ref.~\cite{Pakvasa:2004hu} earlier. 

Assuming the mass degeneracy of $\nu_1$ and $\nu_2$, for the neutrino splitting rates, we can write $\Gamma(\nu_1 \to 3 \nu_3) \approx \Gamma(\nu_2 \to 3 \nu_3)$ for 
inverted mass hierarchy, whereas for normal hierarchy it reads $\Gamma(\nu_3 \to 3 \nu_1) \approx \Gamma(\nu_3 \to 3 \nu_2)$. 
Neutrino mixing angles are also assumed as $\theta_{23} \approx 45^{\circ}, \theta_{13} \approx 0^{\circ}$ and PMNS matrix elements are computed taking these in account,
\begin{align}
 U_{PMNS} = 
\begin{pmatrix}
c_{12} & s_{12} & \epsilon \\
s_{12}/\sqrt{2} &  c_{12}/\sqrt{2} & 1/\sqrt{2} \\
s_{12}/\sqrt{2} &  c_{12}/\sqrt{2}  & 1/\sqrt{2},
\end{pmatrix}
\label{pmns}
\end{align}
where $c_{12}= \cos\theta_{12}, s_{12}= \sin\theta_{12}$, with $\theta_{12} \approx 32^{\circ} $ being the solar neutrino mixing angle, $\epsilon \approx 0.15$. 
The neutrino flux in the mass basis before the decay is 
\begin{align}
\phi_{\nu_i} = \sum\limits_{\alpha} |U_{i \alpha }|^2 \phi_{\nu_{\alpha}}.
\label{mb}
 \end{align}
For normal hierarchy, splitting of $\nu_3$ happens with equal probability to three $\nu_1, \nu_2$ states each.
Assuming that the neutrino decay is complete before it reaches the detector, only stable states $\nu_1, \nu_2$ are included in the final state flux. 
It also ensures that the exponential energy dependent decay factor $e^{-\frac{L m}{E \tau}}$ does not modify the flux. After the splitting, the flux ratio takes the form 
$\phi_1^d : \phi_2^d : \phi_3^d = \phi_1 + \frac{3}{2}\phi_3 : \phi_2 + \frac{3}{2}\phi_3 : 0 $, as the neutrino flux $\phi_3$ splits equally to produce 
three daughter $\nu_1$ or $\nu_2$ from one parent neutrino $\nu_3$. 
Final flux ratio, in the flavor eigenstate takes the form 
\begin{align}
 \Phi_{\nu_{\alpha}} = \sum\limits_{i} |U_{ \alpha i}|^2 \Phi_{i}^d.
\label{fnf}
\end{align}

The three neutrino sources that are considered here consist of neutrino generation from the pion decay, muon decay and neutron decay.
Pions are the primary products expected from hadronic collisions of accelerated protons with ambient matter in AGN's. The  pions decay  to $\mu$ and $\nu_{\mu}$ 
and $\mu$ further decays to one $\nu_e$ and $\nu_{\mu}$ each. 
Here the initial flavor ratio of the flux is $\phi_e : \phi_{\mu} : \phi_{\tau} = 1:2:0$. 
The muon damped sources  the muons from the  $\pi$ decay is absorbed before its decay to provide only one initial $\nu_{\mu}$ with initial flux flavor 
ratio $\phi_e : \phi_{\mu} : \phi_{\tau} = 0:1:0$.
Another source of neutrinos is the neutron decay where one $\nu_e$ is originated in $n \to p e \nu_e$ decay, with initial flavor flux 
ratio $\phi_e : \phi_{\mu} : \phi_{\tau} = 1:0:0$.

For all these three cases initial neutrino flux at mass basis is computed using Eq.~\ref{pmns} and Eq.~\ref{mb}. 
Neutrino flux ratio after the neutrino splitting for the case of normal hierarchy is computed using Eq.~\ref{pmns} and Eq.~\ref{fnf}. 
For normal hierarchy, neutrinos coming from pion decay split like $\nu \to 3 \nu$ to produce flavor ratio of daughter neutrinos as 
$\phi_e : \phi_{\mu} : \phi_{\tau} \approx 2:1:1$. When neutrinos come from muon damped sources with a normal hierarchy, they split to provide final flux ratio in the 
flavor basis as $\phi_e : \phi_{\mu} : \phi_{\tau} \approx 1.8:1:1$. For the neutrinos originating from the neutron decay, their splitting results in the final flux 
ratio $\phi_e : \phi_{\mu} : \phi_{\tau} \approx 3:1:1$.

In all these cases, for inverted mass hierarchy $m_{\nu_1} \approx m_{\nu_2} >>  m_{\nu_3}$ the final neutrino 
mass eigenstates are purely $\nu_3$, which in flavor basis contribute to the neutrino flux as $\phi_{\mu} : \phi_{\tau} \approx 1:1$, with $\nu_e$ flux being negligible. 
Therefore with neutrinos with inverted hierarchy, the flavor ratio of the daughter neutrinos after splitting is $\phi_e : \phi_{\mu} : \phi_{\tau} \approx 0:1:1$ for 
any initial flavor ratio. 

\section{Oblique Parameters: S, T, U}
\label{obp}
The radiative corrections to the gauge boson two point functions ($\Pi_{VV}$) are known as oblique corrections. 
Peskin-Takeuchi oblique parameters manely S,T, U are defined by different combination of those modified two point functions.  
The $S$ parameter measures the running of two point functions of the neutral gauge bosons ($ZZ$, $Z\gamma$ and $\gamma\gamma$) between zero
momentum and the $Z$ pole with $p^2= m_z^2$. 
Therefore, it will be specially sensitive to new degrees of freedom at low scales, particularly below the $m_Z$. 
Thus, it becomes very important in the presence of an ultra light neutral scalars, which is the case for the $\nu$2HDM model with exact $Z_2$ symmetry.  
The $T$ parameter measures the amount of custodial symmetry breaking at zero momentum, i.e. the difference between the $WW$ and the $ZZ$ two point functions at $q^2=0$. 
It usually plays a significant role in constraining the parameter space of particles charged under $SU(2)_L$. 
Splitting the masses of particles inside a doublet breaks custodial symmetry and affects $T$. 
The impact of adding a second Higgs doublet in the electroweak precision tests (EWPT), that is encoded in the
oblique parameters $S$, $T$, and $U$, has been discussed in detail in the literature~\cite{Grimus:2007if, Haber:2010bw}.
%
%
\begin{figure}[tb]
 \begin{center}
   \includegraphics[width=7cm]{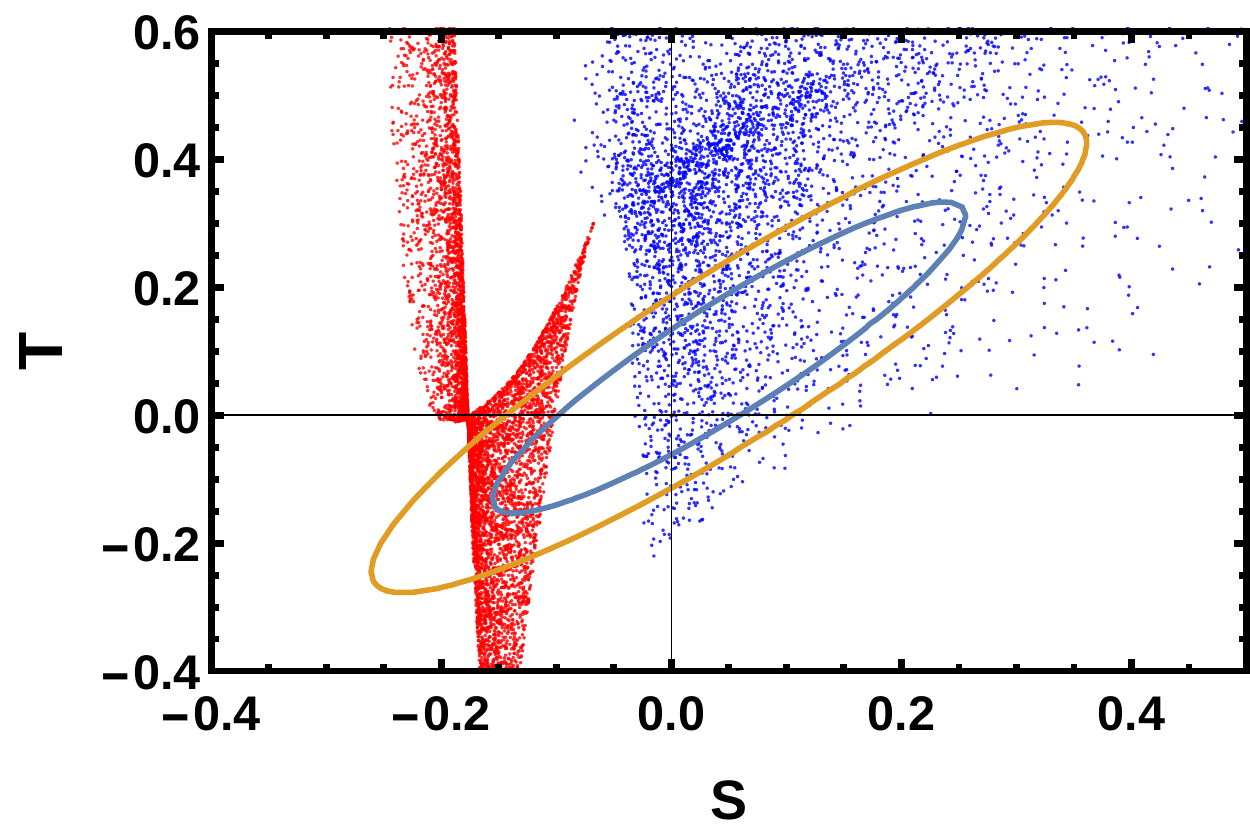} \hspace{0.1cm}
 \end{center}
 \caption{Scatter plot of S and T values in $\nu$2HDM (red points) and in $\nu$2HDM + Vectorlike leptons $\chi, \psi$ (blue points). $m_H=0.1$~eV, $m_{\psi}=m_{\chi}= 150-250$~GeV, $y_{\chi}=1$. Experimentally allowed $1 \sigma$ and $2\sigma$ contours of S, T values are also presented.}
\label{stplot}
\end{figure}
The experimentally measured values of these oblique parameters from the LHC are:
\begin{equation}
  \begin{aligned}
    & \Delta S^{SM} = 0.05\pm0.11,\\
    & \Delta T^{SM} = 0.09\pm0.13,\\
    & \Delta U^{SM} = 0.01\pm0.11,\\
  \end{aligned}
\end{equation}


In the $Z_2$ symmetric $\nu$2HDM, scalar spectrum of the model is very tightly constrained. 
Unbroken $Z_2$ symmetry leads to $m_{12}^2=0$ and the mixing angles are also very small with $\sin^2\alpha,\sin^2\beta\ll 1$. 
While one CP-even scalar $h$ is taken as the 125~GeV Higgs particle discovered at the LHC, and thus has fixed $\lambda_1$. 
The neutrinophilic CP-even neutral scalar $H$ is extremely light, $m_H \sim \text{eV} (O(v_2)) \ll v$. For modest values of the quartic
couplings within the perturbative limit, the charged scalars and the CP-odd scalar masses are expected to be below the TeV scale.
As the presence of a light neutral scalar is essential to explain sufficient depletion of neutrino flux and that kind of scalar is present here, 
it turns out that the oblique parameters S and T will play a decisive role in constraining the model.
The modification of the $S$ parameter due to the presence of an ultra light neutral scalar, $m_H \sim O(v_2)$, pushes it to a large negative value to 
rule out the model at $2 \sigma$ confidence level.

We modify the $Z_2$ symmetric $\nu$2HDM with the addition of one vectorlike lepton $SU(2)_L$ doublet ($\psi$) with hypercharge $Y_{\psi}$ and one 
vectorlike lepton $SU(2)_L$ singlet $\chi$ with hypercharge $Y_{\chi}=Y_{\psi} - \frac{1}{2}$. Both the lepton doublet and singlet are $Z_2$ even. 
So the additional terms that will be added to the $\nu$2HDM Lagrangian being consistent with all the symmetries of the model is
\begin{align}
 L = m_{\psi} \bar{\psi} \psi + m_{\chi} \bar{\chi} \chi - y_{\chi} \bar{\psi} \Phi_1 \chi + h.c. ,  
\end{align} 
After electroweak symmetry breaking $\psi_2$ and $\chi$ mix with each other to provide two lepton mass eigenstates. 
These two along with $\psi_1$ contribute to the S,T parameter calculation. 
The values of these parameters for the cases with and without vectorlike leptons are shown in Fig.~\ref{stplot}. 
Phenomenology of this kind of construction is discussed in detail at Ref.~\cite{Ellis:2014dza}.

\section{Other Constraints on $\nu$2HDM}
\label{oc}
The charged Higgs ($H^{+}$) from 2HDM has interesting phenomenology at the LHC. 
The $p p \to H^{+}H^{-}$ pair production happens at the LHC through the exchange of off-shell $\gamma$ and $Z$. 
This 2HDM production rate remains same for the $\nu$2HDM charged Higgs. 
As explained before, the second scalar doublet here has a very small mixing with the scalars of the SM-like doublet. 
The quark decay modes of the charged scalar are highly suppressed with the mixing factor $v_2/v$. 
So the charged scalar decays mostly to the leptonic channels $H^{\pm} \to l\nu$. 
Corresponding bound on the charged Higgs mass from the LEP in this channel is given as $m_{H^{\pm}} > 80$~GeV.
The charged scalars can contribute to muon and electron g-2 calculation at one loop, but the contribution is negligible due to a 
suppression factor $m_l^4/m_{H^{\pm}}^2$ in the amplitude. 
Unlike a general 2HDM where the two loop contributions can be dominant, charged lepton couplings to $H, A$ are suppressed by a factor $\tan\beta$ in the $\nu$2HDM, 
making $g-2$ constraints insignificant in this scenario. 
The charged scalar loop contribution to $h \to \gamma \gamma$ is of the same sign as the W loop while top loop has a greater contribution with opposite sign. 
Therefore, the new $H^{\pm}$ decreases the diphoton decay amplitude, though the reduced amplitude still satisfies the experimental limit. 

There are strong experimental constraints on the lepton flavor violating processes through the loop diagrams, mainly in the $\mu \to e
\gamma$ channel. For this process, the charged scalar mediated decay branching ratio
reads~\cite{Bertuzzo:2015ada}
\begin{equation}
\mathrm{BR}(\mu \to e \gamma) = \mathrm{BR}( \mu \to e\bar{\nu}\nu) \frac{\alpha_\text{EM}}{192 \pi} |\langle m_{\mu e}^2 \rangle|^2 \rho^2 \, 
\end{equation}
where $\rho = (G_F m_{H^\pm}^2 v_2^2)^{-1}$.
The strongest bound on this lepton flavor violating (LFV) decay comes from
the MEG-2 experiment which gives the upper limit as $\mathrm{BR}(\mu\to e \gamma)<5.7 \times
10^{-13}$~\cite{Adam:2013mnn}, while bounds are weaker for the other LFV decay channels ($\tau \to e \gamma, \tau \to \mu \gamma$),
that are vobtained from the BaBar Collaboration.  
In terms of $\rho$ defined earlier, the $90\%$ confidence level bounds read~\cite{Bertuzzo:2015ada},
\begin{equation}
  \rho \lesssim 1.2 \, \mathrm{eV}^{-2}  \ \ \ \text{for} \ \ \ \ \mu \to e \gamma . 
\end{equation}
The limit on $\rho$ translates into a limit where for $v_2 \lesssim 1$~eV one must have the lower bound as $m_H^{\pm}\gtrsim 250$~GeV.
This is the tightest limit on the $v_2$ and $m_{H^\pm}$ parameter space upto now. 
%
With the sensitivity of the MEG expected to be improved further, the bound on $\rho$ is expected to be improved by about one order of magnitude. 
The limits imposed due to the MEG bound on the $(m_{H^\pm}, v_2)$ plane are shown in Fig.~1 of Ref.~\cite{Bertuzzo:2015ada}. 
During the big bang nucleosynthesis (BBN) era 
the ultra light neutral scalar $H$ cannot stay at equilibrium with thermal bath as $H$ with $m_{\nu} > 0.1$~eV decays strongly through the $H \to 2 \nu$ mode in the $\nu$2HDM. 
Therefore, the ultra light bosonic mode which will otherwise be relativistic, does not contribute to the effctive degrees of freedom $N_{\text{eff}}$ in the BBN era.
Astrophysical constraints on the neutrinophilic 2HDM is discussed in the Ref~\cite{Sher:2011mx}.

\section{Icecube Analysis}
\label{icana}
IceCube is one prominent neutrino telescope which can probe high energetic astrophysical 
neutrino emission from supernovae, gamma ray bursts (GRB), active 
galactic nuclei (AGN) or from other possible new cosmic sources. 
IceCube experiment, after a run of six years, has recorded a total of 82 events combining 
both the track and shower events with neutrino interaction vertex inside the detector. Track events in the IceCube detector arise from the charged current interactions of 
muon neutrinos whereas shower events happen due to charged currents of $\nu_e$ and $\nu_{\tau}$ and neutral currents of all neutrino flavor. These neutrino events observed 
at IceCube are together called 'High Energy Starting Events' (HESE). 
We assume the presence of a single astrophysical neutrino flux along with atmospheric 
components to analyze the observed diffuse neutrino flux. 
Flavor components of isotropic astrophysical neutrino flux can be different for different astrophysical sources, as discussed in section~\ref{nusplit}. 
The flavor ratios of the flux after neutrino spliting are also obtained there. 
If we take only the shower events, the daughter neutrinos are dominantly of electron flavor. 
The all flavor initial astrophysical neutrino spectrum with single component is parametrized as 
\begin{equation}
 \Phi (E_{\nu}) = \Phi_{\text{astro}}^{0} \left(\frac{E_{\nu}}{100 \ \text{TeV}} \right)^{-\gamma}
\label{inif}
\end{equation}
When the astrophysical neutrinos after traveling a distance $L$ from the source with energy $E_{\nu}$ gets splitted with decay lifetime $\tau$, the disappearance 
of those results in the suppression of the neutrino flux. So, just after the decay the neutrino energy spectrum takes the form
\begin{equation}
 \Phi_{decay} (E_{\nu}) = \Phi_{\text{astro}}^{0} \left(\frac{E_{\nu}}{100 \ \text{TeV}} \right)^{-\gamma} e^{-\frac{\beta}{E_{\nu}}},
\label{invde}
\end{equation}
when $\beta$ for a neutrino with mass $m$ is given as 
\begin{equation}
 \beta = L \frac{m}{\tau}.
\end{equation}

\subsection{Splitting of Neutrinos}
\label{vd}
%

\begin{table}[]
\centering
\begin{tabular}{|c|c|c|c|c|}
\hline 
\multicolumn{2}{|c|}{} & \multicolumn{1}{|c|}{No Neutrino Splitting} & \multicolumn{2}{|c|}{With Neutrino Splitting}                                        \\ \hline \hline
 $\Phi_{0}^{ast}$ in $(GeV cm^2 s \ sr)^{-1}$ & $\gamma$ & $\chi^2$  &$\beta$ in GeV &  $\chi^2$                               \\ \hline
 $2 \times 10^{-18} $ &   2.8  & 6.2 & $\ 5 \times 10^4 $   &   4.70                  \\ \hline
 $5 \times 10^{-18}$ &  3  & 65.2 & $ 1.25 \times 10^5 $   &    6.18                \\ \hline
\end{tabular}
\caption{Effects of visible splitting (and the absence of it) of active astrophysical neutrinos on $\chi^2$ in the Icecube.}
\label{tab:nudecay}
\end{table}

In the neutrinophilic two Higgs doublet model ($\nu$2HDM) the active neutrinos do not decay to the invisible final state particles. 
Instead, one heavier neutrino mass eigenstate splits to three lighter neutrino mass eigenstates with a process like $\nu_3 \to 3 \nu_1$, 
creating three daughter neutrinos for each initial neutrinos. 

To account for the general time dependence in the neutrino propagation, we take the case where some of the neutrinos have decayed while rest of them have not till that time.
Part of the neutrino flux that is decayed will be increased three times after the decay here, compared to the case when neutrinos just vanish due to 
their decay to invisible particles. 
The amount of the initial flux that remains unchanged after the decay is of the form 
\begin{equation}
\Phi_{nd} (E_{\nu}) = \Phi_{\text{astro}}^{0} \left(\frac{E_{\nu}}{100 \ \text{TeV}} \right)^{-\gamma} e^{-\frac{\beta}{E_{\nu}}}.
\end{equation}
The daughter neutrinos originating after the decay will have different energies compared to the parent one and we denote that 
energy as $E_d$. We assume all three daughter neutrinos will have same energy i.e. $E_d \approx E_{\nu}/3$ and they will interact 
with the IceCube detector with that energy.
So, the part of the flux that is decayed to three neutrinos each, can be written as 
\begin{equation}
\Phi_{d} (E_{\nu}) = \Phi_{\text{astro}}^{0} \left(\frac{E_{\nu}}{100 \ \text{TeV}} \right)^{-\gamma} \left( 1- e^{-\frac{\beta}{E_{\nu}}} \right).
\end{equation}
For a particular bin of detection energy in the IceCube, the daughter neutrino energy will have a lower bound equal to the detection energy. 
With all these considerations in mind we write the total number of astrophysical neutrinos being observed at a particular energy bin as~\cite{Dey:2016sht}
\begin{equation}
 N_i = 4 \pi \Delta t n_{int} \int^{E_{i+1}}_{E_i} dE \left[ \int^{\infty}_{E} \Phi_{nd} (E_{\nu}) \frac{d \sigma}{dE} (E_{\nu})  d{E}_{\nu}
            + \int^{\infty}_{E} 3 \ \Phi_{d} (E_{\nu}) \frac{d \sigma}{dE} (E_d) d{E}_d \right],
\end{equation}
when $ \frac{d \sigma}{dE} = \left( \frac{d\sigma^{NC}}{dE} + \frac{d\sigma^{CC}}{dE} \right)$ is the total neutral current (NC) and charged current (CC) interaction.
 This represents the effective interaction area per effective interacting nucleon present inside the matter of the IceCube experiment. 
Here $n_{int}$ is the total effective number of nucleon-neutrino interaction points at the IceCube, where $\Delta t$ is the total time 
exposure of the IceCube detector i.e. the duration of IceCube run of 2078 days.
 %

\begin{figure}[tb]
 \begin{center}
   \includegraphics[width=7cm]{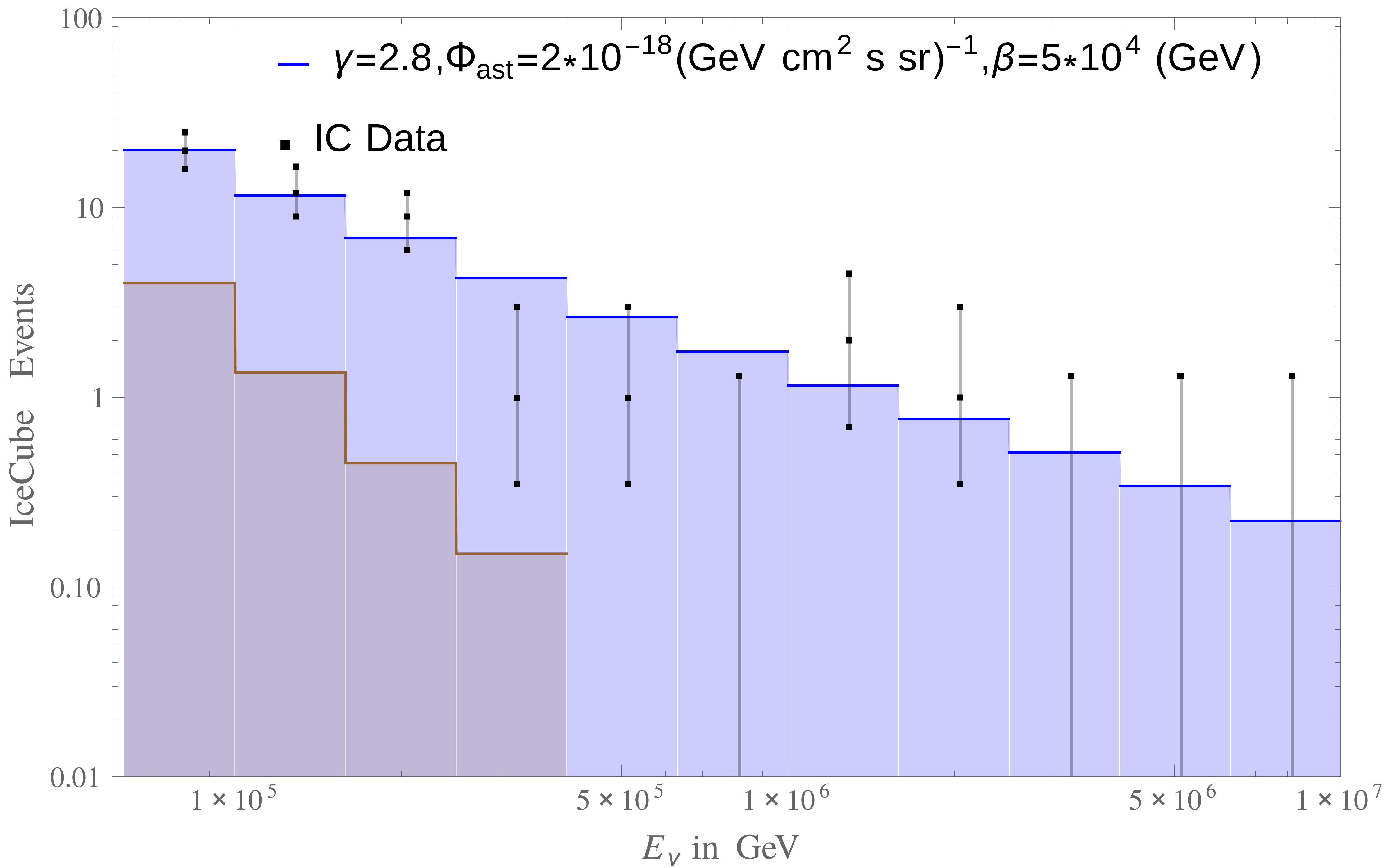} \hspace{0.1cm}
    \includegraphics[width=7cm]{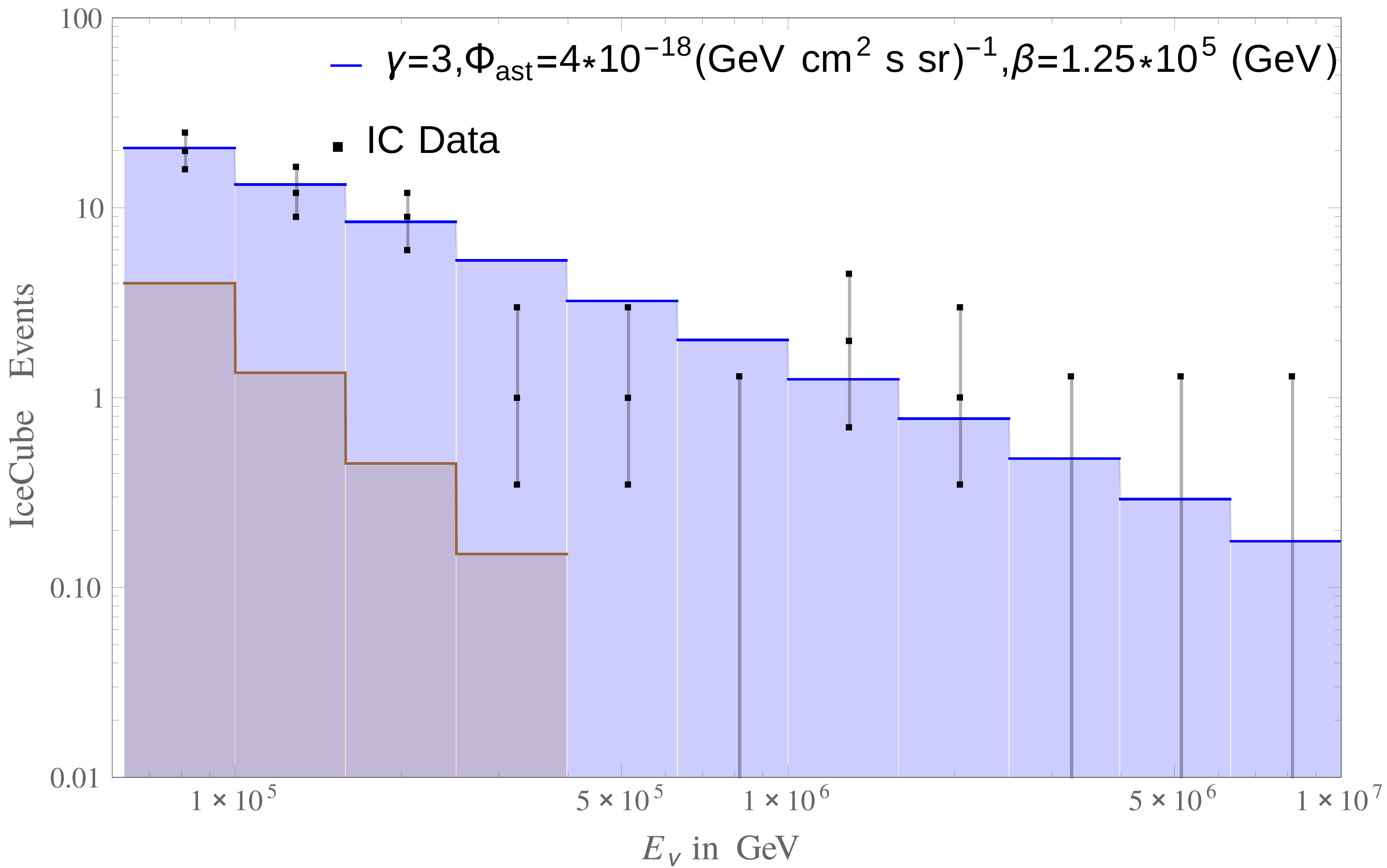}
 \end{center}
 \caption{IceCube event distribution for neutrino splitting with a. moderate flux amplitude, relatively flat spectrum and small decay effects ($\Phi_{astro}^0 = 2 \times 10^{-18} (\rm GeV \ cm^2 s \ sr)^{-1}, \gamma=2.8, \beta = 5 \times 10^4\ \rm GeV$) (left) b. higher flux amplitude, steeper 
spectrum and higher decay effects ($\Phi_{astro}^0 = 4 \times 10^{-18} (\rm GeV \ cm^2 s \ sr)^{-1}, \gamma=3, \beta = 1.25 \times 10^5\ \rm GeV$) (right). Atmospheric neutrino background is shown as brown shaded region.}
\label{decayv}
\end{figure}

The interaction cross-section can be written as~\cite{Dutta:2015dka}, $\frac{d\sigma}{dE} = \frac{1}{E_{d}} \frac{d\sigma (E_d)}{dy}$, where $y$ 
is the energy fraction of final state lepton and can be written here as $\frac{E_l}{E_d}$ and $E$ is the energy deposit. 
A complete shower acceptance simulation, taking into account the difference of energy transfer in CC scattering of $\nu_e$ and $\nu_{\tau}$ along with 
unequal hadronic content in the final states, is beyond the scope of this work. Following the Ref.~\cite{Kistler:2013my}, full transfer of $\nu_e$ energy into showers is 
approximated as, $E_d \approx E$ that can be used to write $d\sigma/dE = \sigma^{CC} (E^e_d) \delta(E-E^e_d)$. Taking into account $50 \%$ energy 
transfer for $\nu_{\tau}$ initiated CC interaction we write $d\sigma/dE = \sigma^{CC} (E^{\tau}_d) \delta(E- 0.5 E^{\tau}_d)$. For all neutrino flavors, 
neutral current interaction is written as $d\sigma/dE = \sigma^{NC} (E_d) \delta(E- 0.2 E_d)$. The energy dependent $\nu_e-$nucleon charged current (CC) and 
neutral current (NC) interaction cross section $\sigma(E_d)$ in the SM case can be extracted from Fig.~3 of Ref.~\cite{Ahlers:2006pf}.   

%
%
\begin{figure}[tb]
 \begin{center}
   \includegraphics[width=7cm]{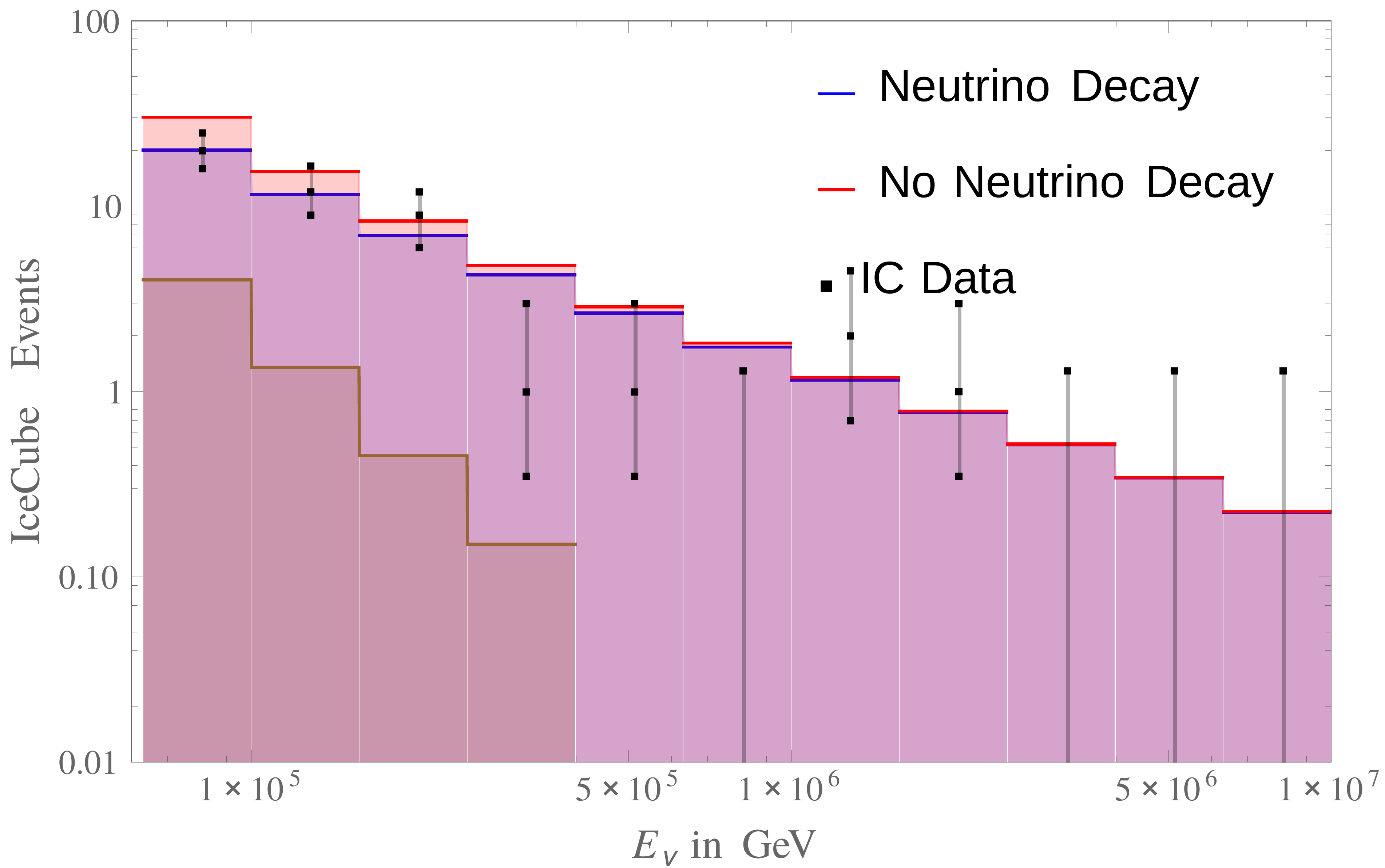} \hspace{0.1cm}
    \includegraphics[width=7cm]{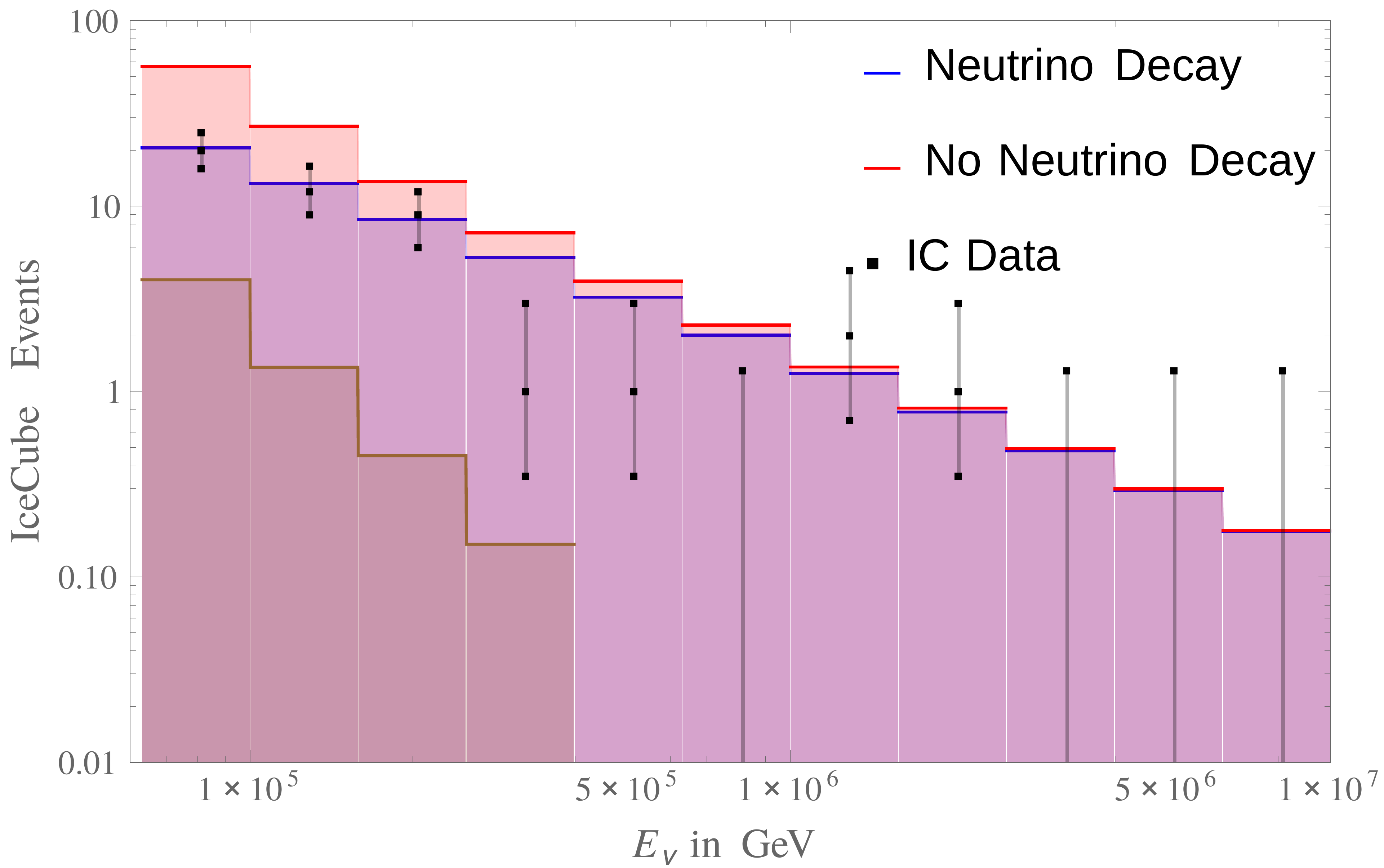}
 \end{center}
 \caption{IceCube event distribution with comparison of the cases of no neutrino decay for a. $\Phi_{astro}^0 =2 \times 10^{-18} (\rm GeV \ cm^2 s \ sr)^{-1}, \gamma=2.8$ 
(left) b. $\Phi_{astro}^0 = 5 \times 10^{-18} (\rm GeV \ cm^2 s \ sr)^{-1}, \gamma=3$ (right) with cases with a visible neutrino 
decay a. $\beta = 5 \times 10^4$\ \rm GeV (left) and b. $\beta = 1.25 \times 10^5$\ \rm GeV (right) respectively. Atmospheric neutrino background is shown as brown shaded region.}
\label{comdnd}
\end{figure}
We choose some benchmark points of the relevant parameters like the astrophysical neutrino flux amplitude $\Phi_{astro}$, the spectral index 
$\gamma$ and $\beta$ that quantifies the amount of neutrino decay. It is shown in Fig.~\ref{decayv} how an active neutrino splitting to three active neutrinos 
can have a significant impact on the shape of the distribution of IceCube events. IceCube event distribution in the energies from 250 
TeV to 1 PeV are better explained if the the neutrino spectrum is relatively steeper. A moderate astrophysical neutrino flux amplitude along with a high spectral
index results in heavy suppression of the number of expected events at energies above 1 PeV causing a large deviation from the observed IceCube neutrino events. 
If the neutrino flux amplitude is increased for a steep spectrum it can explain observed events at energy bins above 250 TeV but the deviation is there 
in the lower energy (60-250 TeV) bins, as the expected number of neutrino events increase with flux amplitude. 
This issue can be resolved introducing a split of active neutrinos and therefore suppressing the energy spectrum with the factor $e^{-\frac{\beta}{E_{\nu}}}$ with 
greater suppression at lower energy bins, which therefore can help expected neutrino events to match the observed ones. The plots in Fig.~\ref{comdnd} present effects of 
the introduction of a splitting of neutrinos for different benchmark points comparing with the case where there is no neutrino splitting. Amount of suppression that 
the decay brings to the neutrino event distribution is dependent on $\beta$, and increases with it. The left plot with $\beta = 3.8 \times 10^5 $GeV suppresses 
the event distribution of the neutrino without decay case more than that in the right plot with $\beta = 2 \times 10^4$~GeV. 

\begin{figure}[tb]
 \begin{center}
   \includegraphics[width=7cm]{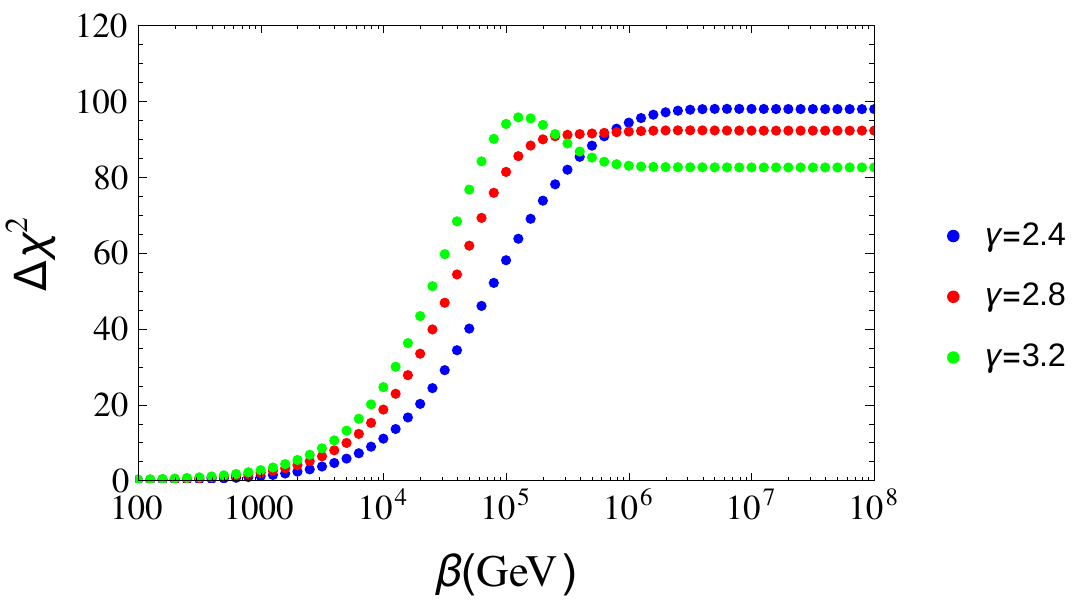} \hspace{0.1cm}
    \includegraphics[width=7cm]{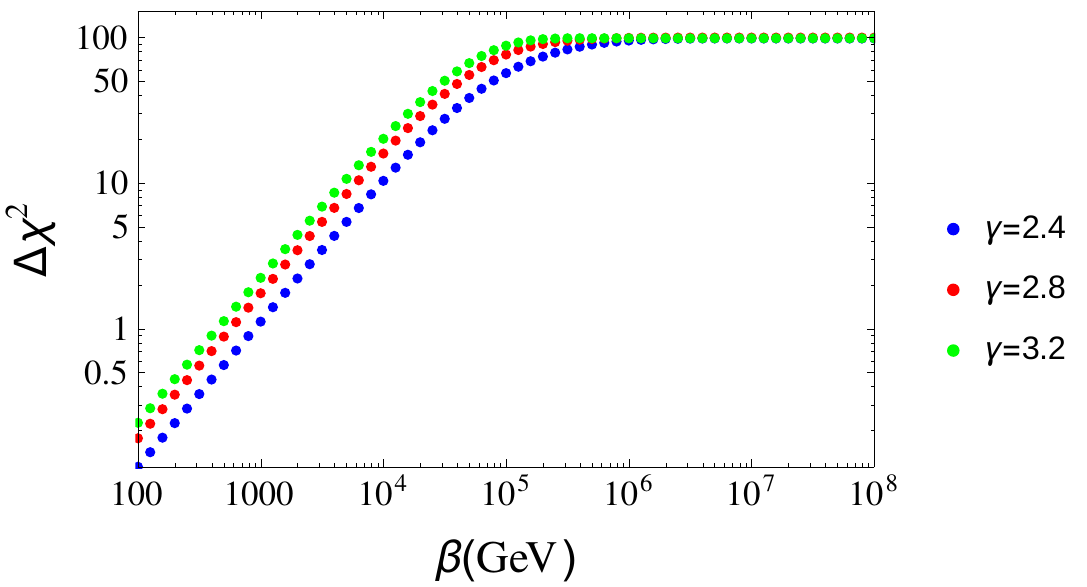}
 \end{center}
 \caption{$\Delta \chi^2$ with $\beta$ for different steepness of the neutrino flux ($\gamma = 2.4, \ 2.8, \ 3.2$) for initial flux amplitude 
a. $\Phi_{astro}^0= 5 \times 10^{-18} (\rm GeV \ cm^2 s \ sr)^{-1}$ (left) b. $\Phi_{astro}^0= 1 \times 10^{-17} (\rm GeV \ cm^2 s \ sr)^{-1}$ (right).}
\label{chisq}
\end{figure}

We quantify the effects of neutrino decay on the $\chi^2$ values in our statistical analysis here. As the $\chi^2$ values by themselves 
are not too informative, we try to compare the $\chi^2$ values with neutrino decay with its values when there way no neutrino decay, over the parameter 
region that is interesting for IceCube phenomenology. Therefore, we use a metric 
that can compare the two cases of absence and presence of neutrino splitting and probe if there 
is a better fit for the case of neutrino decay. We define the quantity as,
\begin{equation}
\Delta \chi^2 = 100 \times \frac{\chi^2(\rm no \ decay) - \chi^2(\rm with \ decay)}{\chi^2(\rm no \ decay)},
\end{equation}
and compute its values with variation of different model parameters.  The variation of $\Delta \chi^2$ over different parameter region is presented in Fig.~\ref{chisq}.

\subsection{Invisible Decay of neutrinos}
\label{invd}
Decay of active neutrinos with absence of any SM neutrinos in the final state decay products is termed as neutrino invisible decay. In the $\nu$2HDM model discussed here,
neutrino decay is not invisible. Invisible decays are possible in different scenarios like lepton number violating triplet Majoron models~\cite{Gelmini:1980re} and 
singlet Majoron models~\cite{Chikashige:1980qk}. Therefore, we like to explore the observable effects of invisible neutrino decay at the IceCube. 
We start with an isotropic astrophysical neutrino flux shown above in Eq.~\ref{inif}.   
The expression of neutrino flux in Eq.~\ref{invde} depicts the suppression of neutrino flux due its invisible decay. 
Using that expression, total number of astrophysical neutrinos being observed at a particular energy bin $[E_i, E_{i+1}]$ as 
\begin{equation}
 N_i = 4 \pi \Delta t n_{int} \int^{E_{i+1}}_{E_i} dE \int^{\infty}_{E}  \Phi_{decay} (E_{\nu}) \left( \frac{d\sigma^{NC}}{dE} + \frac{d\sigma^{CC}}{dE} \right) d{E}_{\nu} ,
\end{equation}
when $\left( \frac{d\sigma^{NC}}{dE} + \frac{d\sigma^{CC}}{dE} \right)$ is the neutrino-nucleon interaction as discussed and computed above. 
\begin{figure}[tb]
 \begin{center}
   \includegraphics[width=7cm]{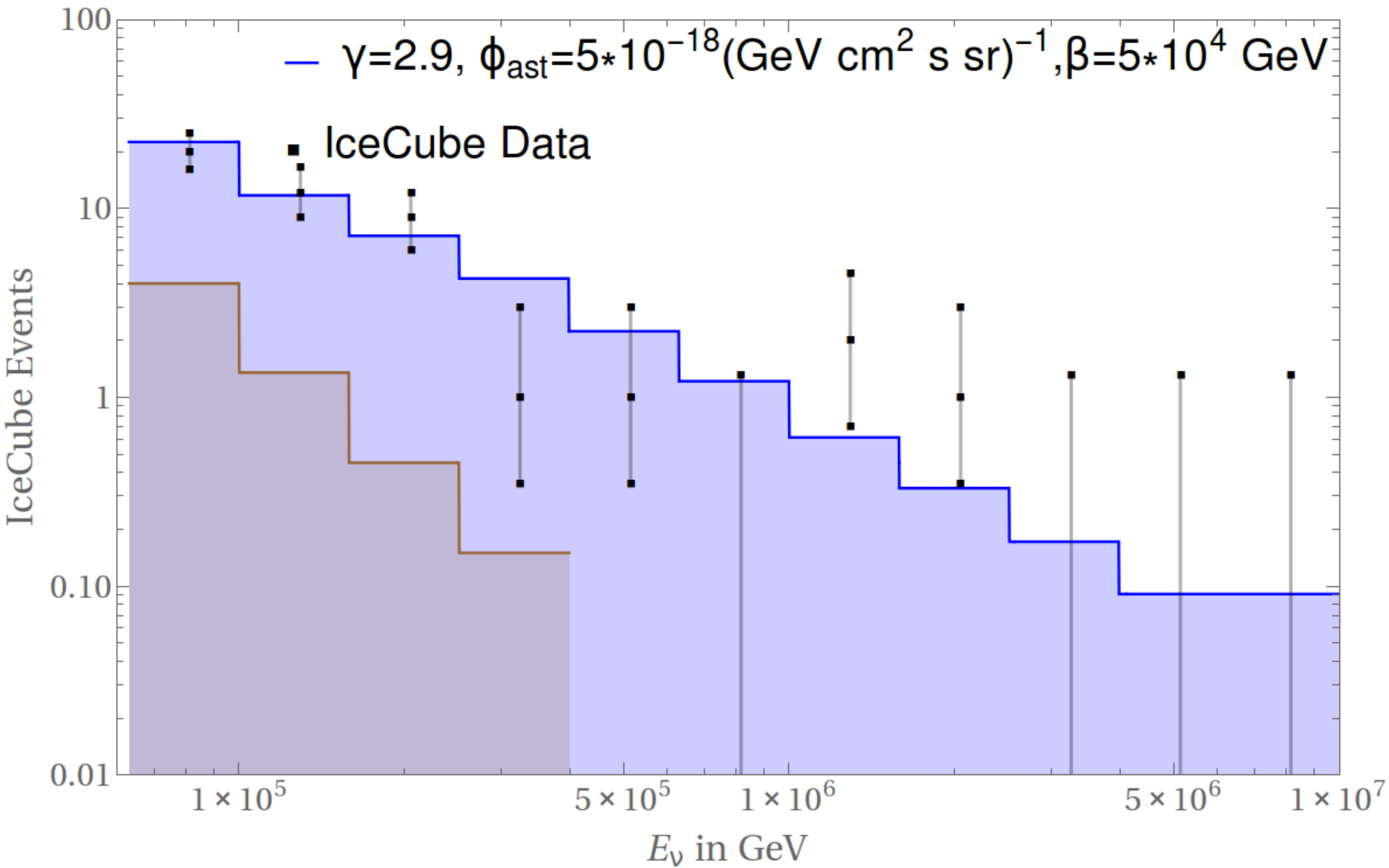} \hspace{0.1cm}
    \includegraphics[width=7cm]{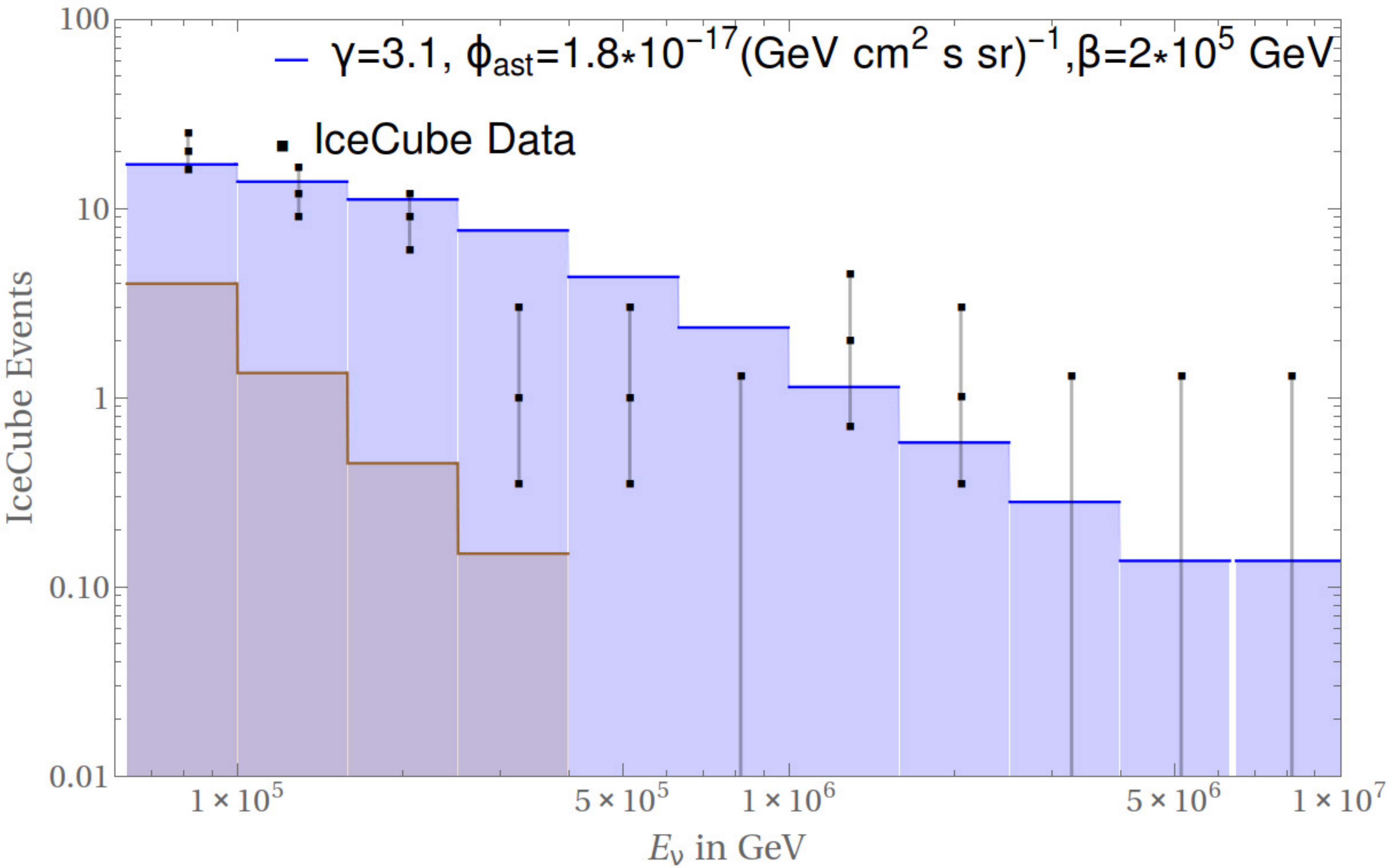}
 \end{center}
 \caption{IceCube event distribution for an invisible neutrino decay with a. moderate flux amplitude, relatively flat spectrum and small decay effects 
($\Phi_{astro}^0 = 8 \times 10^{-17} (\rm GeV \ cm^2 s \ sr)^{-1}, \gamma=3, \beta = 5 \times 10^4 \ \rm GeV $ (left) b. high flux amplitude, steeper spectrum and 
large decay effects ($\Phi_{astro}^0 = 2 \times 10^{-17} (\rm GeV \ cm^2 s \ sr)^{-1}, \gamma=2.75, \beta = 2 \times 10^5 \ \rm GeV$) (right). Atmospheric neutrino 
background is shown as brown shaded region.  
 }
\label{invdecay}
\end{figure}
How the distribution of events are modified at IceCube in presence of an invisible neutrino decay is shown in Fig.~\ref{invdecay}. For a flat energy 
spectrum of astrophysical neutrinos, IceCube events in multi-PeV bins are satisfied along with those in the energies upto 250 TeV. That results in the 
intermediate energy bin event distribution being far off from the expected results. Similarly, a flat energy spectrum which satisfies events in all bins above 250 TeV,
falls short of observed events in the lower energy bin. A relatively steeper neutrino energy spectrum with higher $\gamma$ can satisfy all the energy bin above 250 TeV, 
but with high spectrum amplitude that results in excess events at lower energy (60 -250 TeV) bins. Introduction of an invisible neutrino decay can suppress the flux 
at lower energy to satisfy the IceCube event distribution. 

\section{Summary and Conclusion}
\label{conc}
The anomalies in the IceCube data have been  addressed in terms of excess at PeV energies or a cutoff at 6 PeV Glashow resonance energies. In this paper we discuss neutrino splitting which transfers neutrino energies to roughly 1/3 after decay and due to the Lorentz factor there is a larger depletion of lower energy neutrinos. 
The two effects together give an explanation of the PeV excess and 6 PeV depletion in a single power law initial neutrino flux scenario. The benchmark points that explain the IceCube spectrum put the decay parameter in the range, $\beta=L/(\tau/m)= (5 \times 10^4 - 1.25 \times 10^5)$ GeV with the correlated spectral index being in the range
$\gamma = (2.8-3) $ as shown in Table~\ref{tab:nudecay}. The $\chi^2$ values for benchmark points in the case with no neutrino splitting or decay along with the case when they are present are also shown in Table~\ref{tab:nudecay}. 
To achieve a decay at $ L=1$ Mpc of $m_\nu=0.1 $~eV neutrinos we need $\tau= (3 \times 10^{-2} - 2 \times 10^{-1} )$~second. 
To achieve this value of $\tau/m$ we need a small mass mediator. This small mass scalar occurs naturally in a $\nu$2HDM model where the $vev$ of the scalar is taken to $O(eV)$ in order to explain the neutrino mass. Presence of an ultra light scalar along with the mass splitting between the neutral and charged component of the second Higgs doublet gives rise to a large contribution to the $S $ and $T$ oblique parameters. This problem is fixed here by adding a vector lepton doublet and singlet with mass 200 GeV. The neural component of the heavy fermions can serve as dark matter as has been studied in~\cite{Bhattacharya:2015qpa}. The charged and CP-odd neutral scalars and charged vector-like leptons can potentially be observed at the LHC, along with interesting phenomenological properties~\cite{Kumar:2015tna, Wang:2016vfj, Gopalakrishna:2015wwa}.

For the neutrinos splitting $\nu \rightarrow 3 \nu$ scenario the daughter neutrinos are in the ratio $\phi_e : \phi_{\mu} : \phi_{\tau} \approx 0:1 :1$ for the case of inverted hierarchy and for normal hierarchy the ratio is $\phi_e : \phi_{\mu} : \phi_{\tau} \approx 2:1 : 1$ with pion decay as neutrino source, $\phi_e : \phi_{\mu} : \phi_{\tau} \approx 1.8:1 : 1$ for muon damped source of neutrinos and $\phi_e : \phi_{\mu} : \phi_{\tau} \approx 3: 1 : 1$, when neutrinos come from the neutron decay. So the flavor ratio determination \cite{Aartsen:2015ivb} can be a good test of the neutrino splitting scenario which can also give information of the mass hierarchy of neutrinos and distinguish the neutrino splitting scenario from other phenomenon like invisible decay of neutrinos \cite{Pakvasa:2004hu, Maltoni:2008jr, Baerwald:2012kc}.

\acknowledgments
SS acknowledges the useful discussion with Bhavesh Chauhan on the effects of an ultra light scalar on the BBN constraints. 


\end{document}